\documentclass[useAMS,epsfig,usegraphicx,twocolumn,usenatbib] {mn2e}
\def\aj{AJ}
\def\araa{ARA\&A}
\def\apj{ApJ}
\def\apjl{ApJL}
\def\aap{A\&A}
\def\mnras{MNRAS}
\def\nat{Nature}

\def\physrep{Physics Reports}

\usepackage{graphicx}
\usepackage{amsmath, amssymb}

\usepackage[usenames,dvips]{color}

\newif\ifAMStwofonts

\makeatother

\title[Angular momentum transport ]
{Angular momentum transport and evolution of lopsided galaxies}

\author[Saha \& Jog]
{Kanak Saha$^{1}$\thanks{E-mail:kanak@iucaa.ernet.in}, \& Chanda J. Jog$^{2}$\\
1. Inter-University Centre for Astronomy and Astrophysics, Pune 411007, India\\
2. Department of Physics, Indian Institute of Science, Bangalore 560012, India\\}
 
\begin{document}

\date{Accepted xxxx Month xx. Received xxxx Month xx; in original form \today} 
\pagerange{\pageref{firstpage}--\pageref{lastpage}} \pubyear{2014}
\maketitle

\label{firstpage}

\begin{abstract}
The surface brightness distribution in the majority of stellar galactic discs
falls off exponentially. Often what lies beyond such a stellar disc is the 
neutral hydrogen gas whose distribution also follows a nearly
exponential profile at least for a number of nearby disc galaxies. Both the 
stars and gas are commonly known to host lopsided asymmetry especially in the outer 
parts of a galaxy. The role of such asymmetry in the dynamical evolution of 
a galaxy has not been explored so far.

Following Lindblad's original idea of kinematic density waves, we show that
the outer part of an exponential disc is ideally suitable for hosting 
lopsided asymmetry. Further, we compute the transport of angular momentum 
in the combined stars and gas disc embedded in a dark matter halo.
We show that in a pure star and gas disc, there is a transition point 
where the free precession frequency of a lopsided mode, $\Omega -\kappa $, changes
from retrograde to prograde and this in turn reverses the direction of angular momentum flow
in the disc leading to an unphysical behaviour. We show that this problem is overcome 
in the presence of a dark matter halo, which sets the angular momentum flow 
outwards as required for disc evolution, provided the lopsidedness is leading in nature.
This, plus the well-known angular momentum transport in the inner parts due to spiral arms, 
can facilitate an inflow of gas from outside perhaps through the cosmic filaments.

\end{abstract}


\section{Introduction}
\label{sec:intro}

With the expansion of the universe, the frequency of violent events such
as interaction and mergers are expected to diminish leaving a galaxy
nearly isolated. The evolution of such galaxies is mainly
 dependent on the internal processes which are generally 
slow with a typical time scale of a billion year. These slow internal
processes which are believed to be the dominant mechanism of evolution in 
the later quiescent phase of galaxy evolution, are better known as the cause 
of secular evolution \citep{KormendyKennicut2004}. 
Various non-axisymmetric features such as bars, 
spiral arms ($m=2$ perturbation) are the commonly recognized driver of this evolution.
As shown by the pioneering work of \cite{LBK1972}, spiral arms are 
able to lead the secular evolution via redistributing energy
and angular momentum within the galaxy such that they eventually
manage to enhance the central concentration as observations 
indicate \citep{PfennigerNorman1990,Zhang1999,Laurikainenetal2007,Kormendy2008}.
An important outcome of \cite{LBK1972} paper is that in order to drive 
the evolution in the right direction (i.e., by angular momentum flowing outward),
the spiral arms have to be trailing a fact that was difficult to be determined by
the observations at the time. The role of bars too are now well established in this context.
Bars are known to grow via transferring angular momentum outward and eventually 
goes through the buckling instability to form a boxy/peanut 
bulges \citep{Combesetal1990,Rahaetal1991,Athanamisi2002,Debattistaetal2004,
Sahaetal2010,SahaNaab2013}. Apart from hosting spiral arms and bars, many disc galaxies 
are also lopsided ($m=1$ perturbation) and generally the asymmetry is prominent 
in the outer parts of the galaxy. However, the precise role of lopsided asymmetry 
in driving the evolution in galaxies has been little explored.

Indeed, more than a third of the spiral galaxies host large-scale lopsided 
asymmetry \citep{RichterSancisi1994, rixzaritsky1995, Bournaudetal2005, Zaritskyetal2013}, 
see \cite{JogCombes2009}, for a discussion of the threshold of lopsidedness.
Some of the mechanisms that are shown to generate lopsidedness in 
the outer galaxy are: cooperation of orbital streams to a lopsided 
pattern \citep{EarnLyndenBell1996}; response of an axisymmetric disc to 
a lopsided dark halo \citep{Jog1997,Jog1999}; satellite infall onto a 
galaxy \citep{ZaritskyRix1997}; a tidal encounter \citep{Bournaudetal2005, Mapellietal2008}; 
asymmetric gas accretion \citep{Bournaudetal2005} or through internal 
disc instabilities \citep{Sahaetal2007,Duryetal2008}. No unique mechanism for generating 
lopsidedness has been identified so far, and it could well be that
a combination of different processes is at work. Further, it has been shown 
that for isolated galaxies, the amplitude of lopsidedness is uncorrelated 
with the strength of tidal interaction \citep{Bournaudetal2005,vanEymerenetal2011a} 
or the presence of nearby companions \citep{WilcotsPrescott2004}. 
Whatever might be the origin, both the stellar distribution \citep{Blocketal1994,rixzaritsky1995,
Bournaudetal2005} and the neutral hydrogen gas distribution in the outer parts
\citep{Baldwinetal1980,RichterSancisi1994,Angirasetal2006,Angirasetal2007,vanEymerenetal2011a,vanEymerenetal2011b} of a galaxy are known to be lopsided. The strength of lopsidedness, 
as measured by the fractional Fourier amplitude of the $m=1$ component is, on average, 
comparable to that of the $m=2$ component as seen in spiral structure and 
bars \citep{JogCombes2009}. Despite this, surprisingly, the dynamical effects of 
lopsidedness on galaxy dynamics and evolution have not received much attention. 
Being a prominent asymmetry in the outer parts, lopsidedness can produce torque the 
way bars and spirals do in the inner regions, and redistribute
angular momentum in the galaxy. But the direction and detailed mechanism of this 
angular momentum transport has not been studied yet in detail, and we note that it
will depend crucially on several factors such the underlying mass distribution, 
pattern speed, and overall whether the lopsidedness is trailing or leading.

Determining the sense of rotation in a galaxy has been a long-standing challenge 
\citep[e.g.,][]{PashaTsitsin1979}. The spiral arms ($m=2$) are typically believed 
to be trailing on theoretical grounds because this facilitates outward transport 
of angular momentum \citep{LBK1972} and very recently ALMA shows clear evidence of
a trailing spiral feeding the Seyfert 1 nucleus in NGC 1566 \citep{Combesetal2014}. 
On the other hand, in the case of a lopsided ($m=1$) mode, when 
the sense of rotation can be determined at all, it seems to be leading, that too 
in the inner parts as in M31 \citep{ConsidereAthanassoula1982}. In ESO 297-27, 
the inner single arm leads while an outer three arms are trailing 
\citep{Grouchyetal2008}. There is also the reverse case when the outer two-armed 
pattern leads while the inner one-arm trails as in NGC 4622 \citep{Butaetal2003}.
Thus while there is some evidence for lopsidedness being leading in the inner parts,
it needs to be confirmed whether it is leading or trailing in the outer parts.

The current paper focuses on the issue of angular momentum flow due to
a lopsided perturbation in a disc galaxy. We follow the treatment of 
\cite{LBK1972, GoldreichTremaine1979, BT2008} to derive the total torque which is a sum
of gravity torque and advective torque (also known as Lorry transport as dubbed by
\cite{LBK1972}) imparted by a lopsided perturbation on the outer disc.
We show that angular momentum flow in a pure exponential disc gives rise to
unphysical result, namely that the sign of angular momentum flow changes abruptly at 
a transition point which corresponds to the point where $\Omega -\kappa$, the kinematic
precession frequency of a lopsided mode changes from being retrograde to prograde. However, 
the inclusion of a dark matter halo sets the angular momentum flow in the outward direction 
as necessary for disc evolution, provided lopsidedness is leading in nature.   

The paper is organized in the following way. Section~\ref{sec:model} describes
the primary model for the disc and halo that we consider. Section~\ref{sec:dispersion} 
contains the calculation of the dispersion relation for the lopsided modes.
The basic calculation of torque and hence the angular momentum transport (hereafter AMT) 
due to a lopsided
perturbation are presented in section~\ref{sec:AMT}. Section~\ref{sec:slow} contains 
results and Section~\ref{sec:discuss} contains discussion. 
The primary conclusions drawn from this work are presented in section~\ref{sec:conclusion}.
The appendix contains calculation and handy formula for obtaining different
disc frequencies for various mass distribution.

\section{Disc and halo model}
\label{sec:model}
We consider three components to model a typical nearby disc galaxy namely stars, 
neutral gas and dark matter. For each of these components, analytic density profiles
are employed to obtain the potential and other dynamical parameters. The galactic
cylindrical co-ordinates $R, \varphi, z$ are used throughout.

The density distribution of stars in our model galaxy follows 
an exponential fall-off with central surface density $\Sigma_{s0}$ 
and a scale length $R_d$ as \citep{Freeman1970}

\begin{equation}
\Sigma_{s}(R)= \Sigma_{s0} e^{-R/R_d}.
\end{equation}

Motivated by the findings of \cite{BrigielBlitz2012}, we use the following
exponential distribution as seen in a number of nearby disc galaxies

\begin{equation}
\Sigma_{g}(R)= \Sigma_{g0} e^{- 1.65 \times {R}/R_{25}},
\end{equation}

\noindent where the fitted central gas surface density $\Sigma_{g0}=28.2$~M$_{\odot}$~pc$^{-2}$ 
according to \cite{BrigielBlitz2012} and $R_{25}$ denotes the radius where the B-band surface
brightness drops to $25$~mag~arcsec$^{-2}$. We use a different value for $\Sigma_{g0}$ as 
mentioned below.

\begin{figure}
\rotatebox{270}{\includegraphics[height=8.5cm]{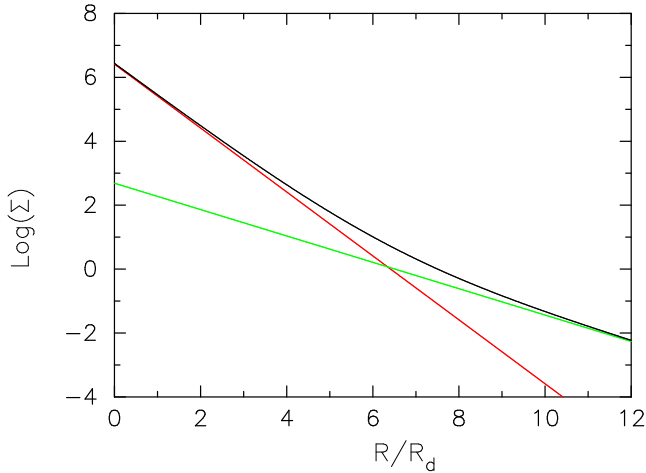}}
\caption{Exponential surface density distribution for the stars and gas. The red curve
denotes stars, green one for the gas. The total surface density is denoted by the solid 
black line. The parameters used are given in section~\ref{sec:model}.
The unit of surface density is in $M_{\odot}pc^{-2}$.}
\label{fig:surfden}
\end{figure}

Then net surface density in the disc is given by 
$$\Sigma_{0}(R)=\Sigma_{s}(R)+\Sigma_{g}(R),$$ 
and as shown in Fig.~\ref{fig:surfden}.
 
\medskip
The expression for the potential and epicyclic frequencies can be found
in \cite{BT2008}. The general expression for an epicyclic frequency in terms of
the potential $\Psi$ is given as

\begin{equation}
{\kappa}^2= \left[\frac {{\partial}^2 \Psi}{\partial R^2}+\frac{3}{R}\frac{{\partial}\Psi}{\partial R} \right]_{z=0}
\end{equation}

\noindent For an exponential disc, the potential at the mid-plane is given as

\begin{equation}
{\Psi_{s}}(R,0)=-{\pi}G {\Sigma_{s0}} R [I_0(y)K_1(y)-I_1(y)K_0(y)],
\end{equation}

\noindent where $y= R/2R_d$; $I_0$, $I_1$ and $K_0$, $K_1$ are the modified 
Bessel functions of first and second kind respectively. Then the epicyclic 
frequency is obtained to be

\begin{equation}
{\kappa}^2_{disc}=\frac{{\pi} G {\Sigma_{s0}}}{R_d} [4 I_{0} K_{0} -2 I_{1} K_{1}+ 2y (I_{1} K_{0}-I_{0} K_{1})]
\end{equation}

\noindent Similarly, the relevant potential and epicyclic frequency for the gas component are 
derived.

The dark matter halos is modelled as an axisymmetric pseudo-isothermal halo with 
a density distribution given by \citep{deZeeuwPfenniger1988}

\begin{equation}
\rho_{dm}(R,z) = \frac{\rho_0}{1 + \frac{S_{a}^2}{R_c^2}},
\end{equation}

\noindent where $S_{a}^2  = R^2 + z^2/q^2$, $q$ denotes flattening or the ratio of the 
vertical to planar axes and $\rho_0$ is the central density and $R_c$ is the core radius. 
The corresponding potential in the spherical co-ordinates is obtained by solving the 
Poisson equation. This is then written in the cylindrical co-ordinates
and from this an expression for $\kappa_{halo}$ is obtained using the general definition 
for the epicyclic frequency given in Eq.~[3].

In order to keep it simple and highlight the primary results from this work, we restrict
ourselves to a particular model for which we have presented all the calculations here.
For the stars, we have used $\Sigma_{s0} = 610.3$ M$_{\odot}$~pc$^{-2}$ 
and $R_d = 3$~kpc -- implying a total stellar mass ($M_d = 2 \pi \Sigma_{0} R_{d}^2$) 
of $3.45\times 10^{10}$~M$_{\odot}$. 

For the gas distribution, we use $\Sigma_{g0}=14.7$~M$_{\odot}$~pc$^{-2}$
and $R_{25} = 4 R_d$ which gives a total gas mass ($M_g$) in the model as
$4.88 \times 10^{9}$~M$_{\odot}$ which is about $14$\% of the stellar mass 
(i.e., $M_g = 0.14 M_d$). The reason for choosing a lower value for the $\Sigma_{g0}$ 
is that we wanted the total gas mass 
to be less than $\sim 20$\% of the chosen total stellar mass so that the disc does not go 
unstable. However, since we are not worrying about the stability of the total disc, 
these values are arbitrary at the moment but agree with the range of typical values seen
 in a galaxy.

\begin{figure}
\rotatebox{270}{\includegraphics[height=8.5 cm]{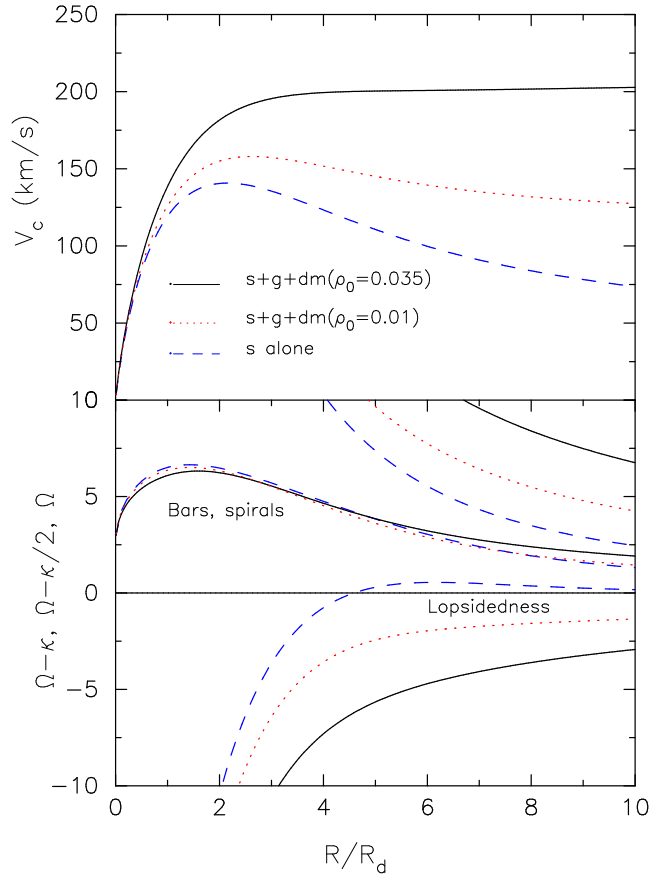}}
\caption{The upper panel showing the total circular velocity curves for two different models of
 stars+gas+dark matter and one with stars alone (shown for comparison). In the lower panel, we show their corresponding frequency profiles: circular frequency $\Omega$ (top 3 curves), $\Omega-\kappa/2$ for m=2 (middle 3 curves) and $\Omega -\kappa$ for the slow m=1 mode (bottom 3 curves). The frequencies are in units of kms$^{-1}$kpc$^{-1}$.}
\label{fig:kinematiclop}
\end{figure}

The flattening and core radius of the dark matter 
halo are fixed at $q =0.9$ (considered to be nearly spherical, for simplicity) 
and $R_c = 1.65 R_d$; 
the central density $\rho_{0} = 0.035$~M$_{\odot}$~pc$^{-3}$. The circular velocity 
curve for this stars+gas+dark matter halo configuration is shown in the upper 
panel (solid black line) of Fig~\ref{fig:kinematiclop}. In order to have a falling 
rotation curve and study its
effect on the kinematic description of lopsidedness, we have made minimal changes in the
halo configuration e.g., decreased $\rho_{0}$ to $0.01$~M$_{\odot}$~pc$^{-3}$ (roughly 
by a factor of $3$); the corresponding
circular velocity is shown by the dotted line in the upper panel of Fig~\ref{fig:kinematiclop}.
The falling rotation curve we have considered (with a gradual fall of $30$~kms$^{-1}$
over $30$~kpc) is reasonable - e.g., our own Galaxy shows this \citep{BrandBlitz1993}
which also helps explaining the high amplitude of HI flaring seen in the outer 
parts \citep{Sahaetal2009}.  
 The lower panel of Fig. 2 will be explained in Section~\ref{sec:dispersion}. 

\section{Kinematic lopsidedness and dispersion relation}
\label{sec:dispersion}

We consider both the stellar and gas disc initially 
axisymmetric and embedded in an axisymmetric oblate dark matter halo
whose density distribution are as mentioned in previous section. 
We treat both the stellar and gas disc in the outer part of a galaxy
as cold self-gravitating fluid disc with velocity dispersion ($\sigma$) being
much less than the local circular velocity ($V_c$) i.e., $\sigma/V_c << 1$.
When such a disc is subject to a lopsided ($m=1$) perturbation of the
form $\sim cos(\varphi - \omega t)$, we can write the density, potential
and velocities in the galactic plane as

\noindent $\Sigma(R,\varphi,t) = \Sigma_{0}(R) + \Sigma^{\prime}(R,\varphi,t)$\\
$\Psi(R,\varphi,t) = \Psi_{0}(R) + \Psi^{\prime}(R,\varphi,t)$ \\
$v_R(R,\varphi,t) =0 + v_{R}^{\prime}(R,\varphi,t)$\\ 
$v_{\varphi}(R,\varphi,t) = R\Omega(R) + v_{\varphi}^{\prime}(R,\varphi,t) ,$\\

\noindent Here the circular frequency $\Omega = V_c/R$ and the perturbed
surface density and potential are connected via the Poisson equation as

\begin{equation}
\nabla^2 \Psi^{\prime} = 4 \pi G \Sigma^{\prime} \delta(z)
\label{eq:poisson}
\end{equation}

\noindent Then the linearized equations of hydrodynamics namely the Euler and continuity 
equations can be written as:
\begin{equation}
\frac{Dv_{R}^{\prime}}{Dt} - 2\Omega v_{\varphi}^{\prime} = -\frac{\partial{\Psi^{\prime}}}{\partial R} -\frac{1}{\Sigma_{0}}\frac{\partial P^{\prime}}{\partial R}
\label{eq:radial}
\end{equation}

\begin{equation}
\frac{Dv_{\varphi}^{\prime}}{Dt} + \frac{\kappa^2}{2\Omega} v_{R}^{\prime} = -{\frac{1}{r}}\frac{\partial{\Psi^{\prime}}}{\partial \varphi} -{\frac{1}{R\Sigma_{0}}}\frac{\partial P^{\prime}}{\partial \varphi}
\label{eq:azimuth}
\end{equation}

\begin{equation}
\frac{D\Sigma^{\prime}}{Dt} + {\frac{1}{R}}\frac{\partial(R\Sigma_{0}v_{R}^{\prime})}{\partial R} + {\frac{\Sigma_{0}}{R}}\frac{\partial{v_{\varphi}^{\prime}}}{\partial \varphi} = 0 
\label{eq:continuity}
\end{equation}

\noindent In the above equations ${D}/{Dt}\equiv {\partial}/{\partial t} + \Omega {\partial}/{\partial \varphi}$ and  $\kappa$ is the epicyclic frequency in the disc. 
For a stellar fluid, the self-consistent derivation of anisotropic stress tensor can be done
only for special cases, because of the closure problem \citep{BT2008}. 
Although \cite{Hunter1979} derived a set of dynamical equations for the anisotropic 
pressure assuming circular motions, a general solution and interpretation remained unclear. 
For our calculation, we consider a polytropic equation of
state, $P=\sigma^2 \Sigma^{\gamma}$, with $\gamma =1$ (also called an isothermal equation
of state)  and $\sigma^2 = (\frac{d P}{d\Sigma})|_{\Sigma_{0}}$ constant throughout 
the disc\citep{BT2008}. This is a good approximation for the gas disc where 
various observations indeed indicate that the gas velocity dispersion remains nearly 
constant with radius and does not depend on the local unperturbed mass 
density \citep{Spitzer1978,Malhotra1995,Lewis1984} but not for the stars which are better 
described by a polytropic index $\gamma = 2$ \citep{Kikuchietal1997}. 
Since lopsidedness is mostly prominent in the outer parts of a galaxy where stellar
density drops down significantly (so does the velocity dispersion as 
$\sigma^2 \propto \Sigma$), we, to a zeroth order approximaton, applied the
same isothermal equation of state to describe both the components. We shall
discuss the impact of this simplified assumption on the angular momentum transport in the
following section.
 
Then writing the perturbed variables in terms of Fourier transform as
$X^{\prime}(R,\varphi,t)=\Re(X_{a}(R) e^{i(\varphi -\omega t)})$ and substituting  them back
in the equations of motion, the perturbed planar velocity field (i.e., radial and azimuthal) can be 
written in a compact form:

\begin{equation}
{\bar{v}}_{a} = A {\bar{\chi}}_{a},
\label{eq:vrvphi}
\end{equation}

\medskip

\noindent with ${\bar{v}}_{a} = \begin{bmatrix} v_{Ra} \\ v_{\varphi a} \end{bmatrix}$ and ${\bar{\chi}}_{a} = \begin{bmatrix} {\frac{d \Psi_{ha}/{d R}}{D_{1}}} \\ { \frac{\Psi_{ha}/R}{D_{1}}}  \end{bmatrix}$ and the matrix $A$ is given by,

\begin{equation}
A  = \begin{bmatrix} i(\omega - \Omega) & -i 2\Omega \\ \frac{\kappa^2}{2 \Omega} & -(\omega - \Omega) \end{bmatrix}
\end{equation}

In the above equations, 
$$D_{1}=-i \det(A) = \kappa^2 - (\omega -\Omega)^2,$$ 

and $$\Psi_{ha} = \Psi_{a} + \sigma^2 \frac{\Sigma_{a}}{\Sigma_{0}}.$$

Then writing the radial part of the perturbed 
quantities as $X_{a}(R) = \Re{X_{a}e^{-i\Phi_{p}(R)}}$, with $\Phi_{p}$ being the phase 
and using a WKB approximation \citep{BT2008}, we have the following dispersion 
relation for $m=1$ lopsided perturbation

\begin{equation}
D_{1} - 2 \pi G \Sigma_{0}(R) |k| + \sigma^2 k^2 =0,
\label{eq:dispersion}
\end{equation}

\noindent where the wavenumber $ k = - {d {\Phi_{p}(R)}}/{d R} $.
 
For trailing perturbation $d {\Phi_{p}(R)}/{d R} < 0$, while it is positive for
a leading case. Note that the general form of the dispersion relation remains
unchanged for any fluid described by a polytropic equation of state except the
fact that the velocity dispersion depends on the local surface density and 
polytropic index (as discussed above). The dispersion relation will be useful 
later in deriving angular momentum transport under the WKB limit.
In absence of self-gravity and pressure of the perturbation, 
the slow m=1 lopsided pattern
would precess with a frequency $\Omega_{pf} = \Omega -\kappa$. 
In Fig.~\ref{fig:kinematiclop}, we show the free precession frequencies
for $m=1$ and $m=2$ perturbations in a cold exponential disc. It was Lindblad  who 
first showed \citep[e.g.,][]{BT1987} that for an $m=2$ spiral arm to survive in a disc with strong
differential shear, the radial variation of $\Omega - \kappa/2$ should be
nearly zero and this condition is nearly satisfied in the region around the 
peak of this curve (see Fig.~\ref{fig:kinematiclop}). In fact,
in most spiral galaxies, the spiral arms are contained within the inner parts 
of the optical radius or about two disc scale lengths, taking the optical disc 
to be about $4 - 5$ disc scale lengths \citep{BinneyMerrifield1998}.

We employ a similar analogy to suggest that lopsidedness should be preferentially seen
in the outer parts of a galaxy. For the exponential disc, the differential
precession i.e., the value of radial derivative of $\Omega-\kappa$, rapidly increases
below $4$ ~scale lengths. However, beyond this radius, the radial variation
of $\Omega-\kappa$ flattens out and becomes almost close to zero. This turns out
to be an ideal situation for the $m=1$ lopsidedness to naturally avoid
the differential shear. In general, a naturally favourable situation
for the lopsidedness to survive would be where the circular velocity
falls as one goes outwards. This occurs in a Keplerian case as in a 
disc around the central black hole, this case will be studied in a future paper.

But lopsidedness is observed in galaxies with a flat
rotation curve too. For a flat rotation curve,
$\kappa = \sqrt{2} \Omega$ and the pattern frequency becomes 
$\Omega_{pf} = -0.414 \Omega$ in which case, the differential shear
becomes non-zero. Then inclusion of self-gravity helps as shown previously
by \cite{Sahaetal2007}. Basically, self-gravity tries to reduce 
the radial epicyclic frequency $\kappa$ so that there is a net reduction in the 
radial variation of $\Omega_{pf}$.
The situation gets more favourable when the net circular velocity falls off 
moderately in the outer parts and one such case is shown in 
Fig.~\ref{fig:kinematiclop} in which case the differential precession is much lower
compared to the galaxies with flat rotation curve. 
However, even with the inclusion of global self-gravity, the preferred
region for lopsidedness seems to be the outer parts of a disc, as the
differential shear in the inner region is too strong \citep{Sahaetal2007}.
 There is another physical reason for lopsidedness to be seen 
mainly in the outer region since the self-gravity of the disc resists any distortion 
of type $m=1$ inside of about 2 disc scale lengths (Jog 1999, Jog 2000) as shown 
by a self-consistent disc response to an imposed potential. 

The exponential disc by itself can support kinematical lopsidedness for a long time 
as seen from the near constancy of $\Omega - \kappa$ (Fig.~\ref{fig:kinematiclop}). 
While addition of the  dark matter halo changes this constancy such a way that the disc plus 
halo system is not so supportive of the $m=1$ mode (unless the halo is such that the overall
rotation curve is falling in nature). On the other hand, the inclusion 
of the dark matter halo makes it possible to allow a smooth outward transport of angular 
momentum without any radial break and hence is a dynamically preferred state as shown
in the subsequent section.

\section{Angular momentum transport due to a lopsided pattern}
\label{sec:AMT}

In the following, we compute the torque due to a lopsided perturbation imposed
onto the disc. We assume that the lopsided perturbation is stationary and
not growing. The disc has no net inflow/outflow of matter from outside/inside 
i.e., mass conservation is strictly followed. In addition, we assume that the
dark matter particles do not take part in the angular momentum transport i.e.,
these particles neither gain nor lose angular momentum; the halo is considered here 
as a simple potential bath. 

\begin{figure}
\rotatebox{270}{\includegraphics[height=8.5 cm]{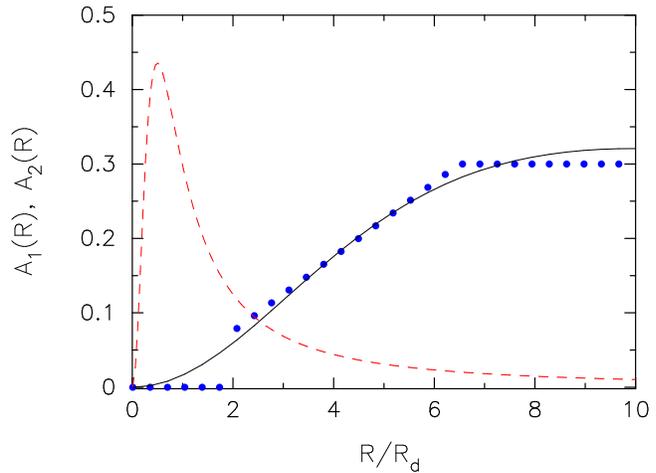}}
\caption{Model radial profiles for the $m=1$ lopsided perturbation $A_1 (R)$ 
and the $m=2$ perturbation, $A_2 (R)$. The filled blue circles
and solid line represent model~1 and model~2 for the lopsidedness (see Eq.~\ref{eq:A1}). 
Whereas, the dashed red line for $A_2$ as given in Eq.~\ref{eq:A2}.}
\label{fig:modelA1}
\end{figure}

\subsection{Gravity transport}
\label{sec:gravity}

Following \cite{LBK1972}, we write the torque exerted by an $m=1$ lopsided perturbation
on the disc material outside a given radius $R_{a}$ as

\begin{equation}
{\mathcal C}_{grav}(R_{a}) \: = \: {\frac{1}{4\pi G}}\int_{-\infty}^{\infty} \int_{0}^{2\pi} R \frac{\partial{\Psi^{\prime}}}{\partial R}  \left. \frac{1}{R} \frac{\partial{\Psi^{\prime}}}{\partial \varphi} \right |_{R=R_{a}} R d\varphi dz.
\end{equation}

\noindent Note that the above integrand is second order in perturbed quantities. There is no 
torque on the disc to first order. 

\noindent We evaluate the gravitational torque in the WKB limit (using 
tightly wrapped potential perturbation) with $\Psi^{\prime}$ given by 
$$ \Psi^{\prime}(R,\varphi,0) = \Psi_{a}(R) e^{i[\varphi - \Phi_{p}(R)]}$$.

Then the resulting expression for the torque is given by

\begin{equation}
{\mathcal C}_{grav}(R_{a}) \: =\: {\frac{1}{4}}{\frac{ R_{a} {|\Psi_{a}(R_{a})|}^{2}}{G}}{\frac{k}{\left |{k^{\prime}}\right |}},
\label{eq:Cgrav}
\end{equation}

\noindent Where $$ k^{\prime} \:=\: \sqrt{k^2 + \frac{1}{R_{a}^2}} $$

\noindent At large distances from the centre of a disc i.e., in 
the outer parts $k^{\prime}\simeq k$. Note that the gravity transport of 
angular momentum depends primarily on the square of the 
perturbing potential and sign of the phase variation of the perturbation.
As is well known, only the trailing spiral ($k>0$) (or any m-fold non-axisymmetric 
perturbation) can exert positive torque on the outer parts and transport
angular momentum outward. The same is, of course, true for the m=1 lopsided
perturbation i.e., the gravity torque due to a trailing lopsided perturbation 
induces outward transport of angular momentum. 
However, gravity torque alone does not tell the full story of angular momentum flow 
in galaxies. Below we discuss another mechanism of angular momentum transport 
proposed by \cite{LBK1972}.

\subsection{Lorry transport}
\label{sec:lorry}

It has been shown that in the presence of a steady non-axisymmetric perturbation, stars
away from resonances neither gain nor lose angular momentum. But these stars nevertheless
help transporting angular momentum just as a system of lorries - a mechanism worked
out by \cite{LBK1972} who named it the "Lorry transport". According to them, in the 
presence of a trailing perturbation, stars with very small eccentricities gain angular
momentum near their apocenters and lose it all near their pericenters. In this sense,
lorry transport of angular momentum by such stars opposes gravity transport of angular
momentum. However, the net advective flux of angular momentum can have either sign when 
derived over a distribution of stars. Now, the total advective torque due to a lopsided 
perturbation can be calculated either following the original derivation of \cite{LBK1972} or
using the fluid equations following \cite{GoldreichTremaine1979, BT2008} which we intend
to use in the present paper. In this section, we calculate the advective torque in both
ways and show under what condition they are equivalent. 

\begin{figure*}
\rotatebox{0}{\includegraphics[height=5.5 cm]{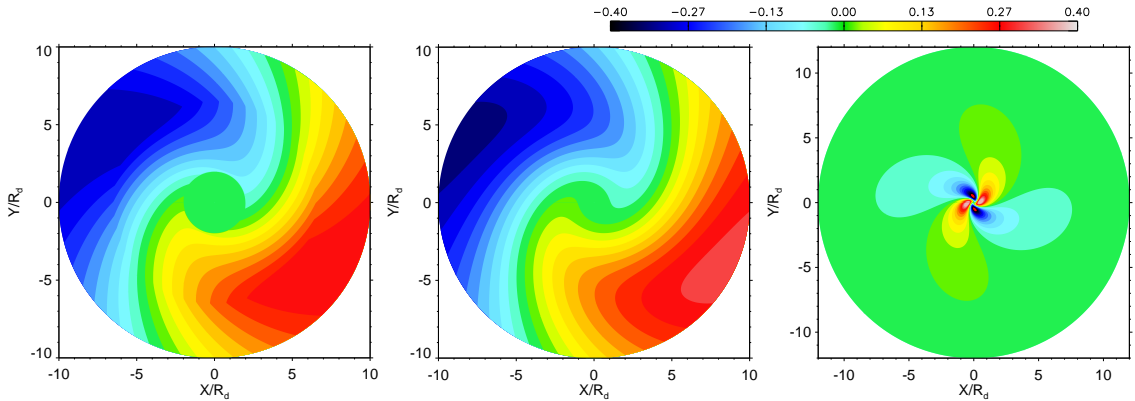}}
\caption{Perturbed surface density maps for the chosen models. The left and middle panels 
represent $m=1$ perturbation corresponding to model~1 and model~2 
respectively (see Eq.~\ref{eq:A1}).
The panel on the right for $m=2$ perturbation for which the radial profiles
are given by Eq.~\ref{eq:A2}. Numerical values for the used parameters are given in 
section~\ref{sec:netAM}. All three panels are scaled to the same color bar.}
\label{fig:surfA1A2}
\end{figure*}

Considering a simple two-integral distribution function (DF), the advective torque due to
 $m=1$ perturbation can be written as \citep{LBK1972}:

\begin{equation}
C_{advec}^{LBK} = -\frac{\pi}{2} \frac{R_{a} \Sigma_{0} {|\Psi_{a}|}^2}{(\kappa a_{0})^2} \frac{\partial}{\partial k} \sum_{l=-\infty}^{\infty} \frac{l \kappa I_{l}({k^{\prime}}^2 a_{0}^2)}{l\kappa + \Omega - \Omega_p},
\end{equation} 

\noindent where $R_a$ is a given radius within which we calculate the torque, $a_{0}$ is 
the average epicyclic amplitude with $J_{1} = \kappa a_{0}^2$ 
denoting the average epicyclic energy.
Then writing $\nu = (\Omega -\Omega_p)/\kappa$, where $\Omega_p$ is the pattern speed 
of a lopsided mode and using the properties of modified Bessel function, we have

\begin{equation}
C_{advec}^{LBK}= -\frac{2 \pi \Sigma_{0} k R_{a} {|\Psi_{a}|}^2}{\kappa^2} \sum_{l=1}^{\infty} \frac{I_{l-1}({k^{\prime}}^2 a_{0}^2) - l/{k^{\prime}}^2 a_{0}^2 I_{l}({k^{\prime}}^2 a_{0}^2) }{1 - \nu^2/l^2}.
\end{equation}

In the longwave limit $k^{\prime} a_{0} < 1$, the dominant contribution comes from the 
$l=1$ term and in that case it can be shown that the advective torque is given by a 
rather simple formula:

\begin{equation}
C_{advec}^{LBK} = - \frac{R_{a}{|\Psi_{a}|}^2 }{G} \frac{\pi G \Sigma_{0} k}{\kappa^2} \frac{1}{1-\nu^2}.
\label{eq:lorryLBK}
\end{equation}

Note that the original derivation for the advective torque by \citep{LBK1972} has 
a typographical error;  in that the term $1 - \nu^2$ should be in the denominator 
as it is in eq.~\ref{eq:lorryLBK} (Lynden-Bell 2014, private communication).

\medskip
Now, we shall derive the advective torque under the fluid approximation. We begin by writing
 
\begin{equation}
C_{advec}^{FL}(R_{a})= R_{a}^2 \int_{0}^{2\pi}{d\varphi \Sigma_{0} v_{\varphi}^{\prime} v_{R}^{\prime}}
\end{equation}

Note that the net torque is again second order in perturbed quantities.
Substituting the perturbed velocities and carrying out the $\varphi$-integral, we have

\begin{equation}
C_{advec}^{FL}(R_{a}) = \frac{\pi}{2}\Sigma_{0} R_{a}^2 (v_{Ra} v_{\varphi}^{\dagger} + v_{Ra}^{\dagger} v_{\varphi}),
\end{equation} 

\noindent $\dagger$ sign indicates the complex conjugate. Substituting 
$v_{Ra}$ and $v_{\varphi a}$ from Eq.~\ref{eq:vrvphi}, we obtain

\begin{equation}
C_{advec}^{FL}(R_{a}) = - {\frac{\pi \Sigma_{0} k R_{a}}{D_{1}}} \Psi_{ha}^2,
\label{eq:Cadvec1}
\end{equation}

\noindent where $D_1$ is defined in section~\ref{sec:dispersion}. Further using the WKB 
solution for a tightly wrapped lopsided density perturbation,

\begin{equation}
\Psi_{ha} = ( 1 - \frac{\sigma^2 |k|}{2 \pi G \Sigma_{0}} )\Psi_{a}
\end{equation}

Substituting the above in eq.~\ref{eq:Cadvec1} and using $D_{1}=\kappa^2 (1 - \nu^2)$ 
which we shall use interchangibly, it is possible to write the advective torque in a 
more familiar form:

\begin{equation}
C_{advec}^{FL}(R_{a}) =  - \frac{R_{a}{|\Psi_{a}|}^2 }{G} \frac{\pi G \Sigma_{0} k}{\kappa^2} \frac{( 1 - \frac{\sigma^2 |k|}{2 \pi G \Sigma_{0}} )^2} {1-\nu^2}
\label{eq:CadvecFL2}
\end{equation}

Note that in the limit $\sigma \rightarrow 0$, we have 

\begin{equation}
C_{advec}^{FL}(R_{a}) = C_{advec}^{LBK}(R_{a})
\end{equation}

The above equality which holds in the limit $\sigma \rightarrow 0$ implies that 
we basically have a pressure-less stellar fluid. The two-integral DF assumed 
by \cite{LBK1972} also amounts to saying that the stars in epicyclic motion 
form a  pressure-less cold self-gravitating system. 
From now on, we shall use eq.~\ref{eq:CadvecFL2} to further deduce the value of advective
torque. Utilizing the dispersion relation given by eq.~\ref{eq:dispersion}, we can write
eq.~\ref{eq:CadvecFL2} as

\begin{equation}
C_{advec}^{FL}(R_{a}) =  - sign(k) \frac{R_{a}{|\Psi_{a}|}^2 }{4 G} \frac{\kappa^2 (1 - \nu^2)}{\pi G \Sigma_{0} |k|}
\label{eq:CadvecFL3}
\end{equation}    

Eq.~\ref{eq:CadvecFL3} is worth examining. Apart from the wavelength
of the perturbation, it depends on the underlying mass distribution and the perturbing potential.
In general, for a disc galaxy we have the inequality 
$\Omega \le \kappa \le 2 \Omega$ satisfied over a wide range of potential variation
from $\Psi_0 \sim 1/r$ to $\Psi_0 \sim r^2$ respectively. 
Then it follows in a straightforward way that for a general galactic potential, in the presence of
any m-fold slowly rotating non-axisymmetric perturbation with $m \ge 2$, $D_{m} < 0$ within 
the disc, where $D_{m} = \kappa^2 - m^2 (\Omega -\Omega_p)^2$.
This holds true as well for a wide range of self-gravitating astrophysical discs as 
$\kappa^2 < m^2 \Omega^2$ always with $m \ge 2$.
However, the scenario changes for $m=1$; in which case, one could, in principle,
have either situation i.e., $\kappa^2 > \Omega^2$ or $\kappa^2 < \Omega^2$.
As we discuss in detail in Section 5, the exponential discs in galaxies happen to manifest
this intriguing situation. This has far-reaching consequences for the dynamical evolution 
of galaxies as shown here.

\subsection{Net AM transport}
\label{sec:netAM}

The net rate of change of angular momentum within the disc is obtained by 
combining the contribution from both the gravity torque and advective torque as

\begin{equation}
\frac{d L_z}{d t}(R_{a}) = - sign(k) \frac{R_{a}{|\Psi_{a}|}^2 }{4 G}\left[ {\frac{|k|}{\left |{k^{\prime}}\right |}} - \frac{\kappa^2 (1 - \nu^2)}{\pi G \Sigma_{0} |k|}\right]
\end{equation}

Substituting the perturbing potential by the perturbed surface density and using the
fact that $\Sigma_{a}(R)=\Sigma_{0}(R)\: A_{1}(R)$, where $A_{1}(R)$ denotes the radial
variation of lopsidedness in the disc (can be obtained via Fourier decomposition), 
we can write:

\begin{multline}
\frac{d L_z}{d t}(R_{a}) \simeq - sign(k)\: (\pi^2 G {\Sigma_{0}}^2 R_{a}^3) \\
\times \tan^2{\alpha} \: {A_{1}}^2(R_{a}) \left[\cos{\alpha} - \frac{ {D_{1}/\Omega^2} \:{R_{a}\:\tan{\alpha}}}{{\pi G \Sigma_{0}}/\Omega^2}\right]
\label{eq:Lzdot}
\end{multline}
    
Eq.~\ref{eq:Lzdot} is our working formula to compute the angular momentum transport. 
Apparently, neither gravity torque nor advective torque explicitly depends on the velocity 
dispersion of the system as we have used the dispersion relation given by 
Eq.~\ref{eq:dispersion} to remove the explicit dependencies. However, the torque depends
on the wavelength of the perturbation ($2\pi/|k|$) or the pitch angle ($\alpha$) which are, 
of course, related to the velocity dispersion of the system. It is trivial to see that
the net magnitude of torque will change by changing the pitch angle of the lopsided 
perturbation. But it is not clear what is the pitch angle for any lopsided perturbation; it 
does measure how tightly wrapped a lopsided perturbation is. For simplicity, we shall use 
$\alpha \simeq 45\,^{\circ}$ corresponding to a WKB limit of $|kR|\simeq 1$ which crudely
represents a logarithmic variation of the phase angle. Although, this is really pushing
the WKB limit to a stage where it is not valid any more, as is often said, it might 
nevertheless give a good insight into the problem. 

In order to proceed further, we need to
know the functional form of $A_{1}(r)$ in a galaxy. In principle, this should be derived 
self-consistently as one of the eigen modes of the set of hydrodynamic equations 
(Eq.~\ref{eq:poisson} - Eq.~\ref{eq:continuity}) stated in section~\ref{sec:dispersion} 
- similar to the eigen modes (which broadly capture the radially increasing 
behaviour of lopsidedness) of \cite{Sahaetal2007}. Since the amplitude of a linear
eigenmode is arbitrary, we decided to connect our angular momentum transport to 
observed lopsidedness in disc galaxies.
Guided by various observations \citep{rixzaritsky1995,vanEymerenetal2011b}, we use 
the following radial profiles for the lopsidedness in the disc:

\begin{equation}
\begin{split}
A_{1} (R) & = \epsilon_1 (2 \frac{R}{R_d} -1) \: \: \: 2. \le R/R_d \le 6.5 \: \: \: \: (model 1)\\
        & = A_{10} \frac{(R/R_d)^2}{(1 + \beta_{1} (R/R_d)^2)^{\gamma}}, \: \: \: \: (model 2) 
\end{split}
\label{eq:A1}
\end{equation}

\noindent $\epsilon_1$ is the halo perturbation parameter and the linearly rising part of $A_1$
arises as a result of self-consistent response of the disc to halo perturbation \cite{Jog2000} 
-- this we refer to as {model~1}, the first formula in Eq.~\ref{eq:A1}. The radial profile
of $A_1$ corresponding to model~1 uses $\epsilon_1 = 0.025$ and is shown in 
Fig.~\ref{fig:modelA1}.
The other functional form for $A_1$ (smooth radial variation) is chosen ad-hoc but 
motivated by typical observational behaviour of lopsidedness in galaxies -- refer to 
as {model~2}, the second formula in Eq.~\ref{eq:A1}. For model~2, we choose 
$A_{10} =0.0166$, $\beta_{1} =0.02$ and $\gamma=1.5$, so that the average value
 of $A_{1}$ in the range $4 - 6 R_d$ are around $0.25$ (see Fig.~\ref{fig:modelA1}).
By carefully choosing the free parameters, the same functional form can roughly capture
the typical radial variation of $m=2$ perturbation such as bars/spirals \citep[e.g., see][for radial variation of $A_2$]{Durbalaetal2009} 
and we write this as:  

\begin{equation}
A_{2} (R) = A_{20} \frac{(R/R_d)^2}{(1 + \beta_{2} (R/R_d)^2)^{\gamma}}, 
\label{eq:A2}
\end{equation}

\noindent where the free parameters $A_{20}=7.5$, $\beta_{2}=5.0$ and $\gamma=1.8$ gives
rise to a radial profile for $A_2$ as shown in Fig.~\ref{fig:modelA1}.
The perturbed surface density maps corresponding to these $A_1$ and $A_2$ profiles are 
as shown in Fig.~\ref{fig:surfA1A2}. One can, in principle, change these free parameters
to model any other profiles for $A_1$ and $A_2$. 
In the following, we shall employ these profiles for 
the calculation of total torque in the disc exerted by the lopsided perturbation and
the $m=2$ perturbation.  

\begin{figure}
\rotatebox{270}{\includegraphics[height=9.0 cm]{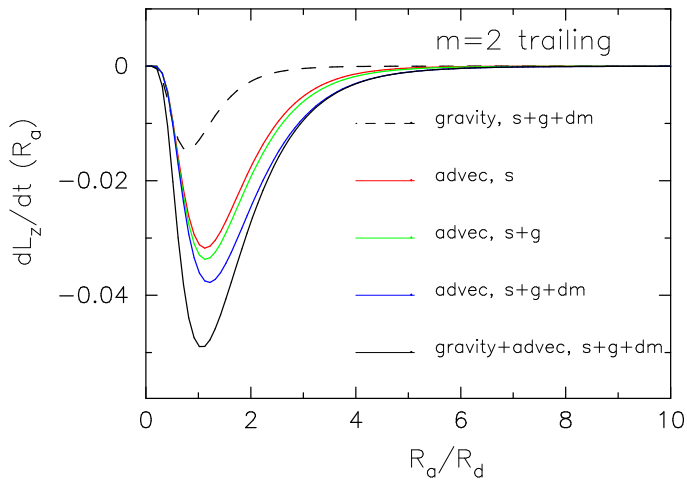}}
\caption{Radial variation of angular momentum flow in the disc for 
$m=2$ trailing spiral. In the figure, 's' stands for stars, 'g' for gas and 'dm' 
for dark matter. Negative~sign on $dL_z/dt (R_{a})$ indicates that the disc within $R_{a}$
looses angular momentum in the presence of an $m=2$ trailing spiral. The unit of $dL_z/dt$
is $G M_d^2/R_d$.}
\label{fig:Lzdotm2}
\end{figure}

From eq.~\ref{eq:Lzdot}, it is obvious that in the absence of an advective torque, any m-fold
trailing non-axisymmetric perturbation would transport angular momentum outward by 
gravity torque alone. When advective torque is added, the net sign of the angular momentum flow
 is not clear apriori; as mentioned elsewhere it can in some cases add on to the effect of 
the gravity torque or oppose it.
One of the purposes of this paper is to explore the parameter space and 
various disc-halo configurations to understand the flow of angular momentum
in disc galaxies - which, in turn, is crucial for understanding the evolution
of galaxies. In the following section, we restrict ourselves to a slowly rotating 
lopsided perturbation and discuss briefly here the $m=2$ case just for the purpose 
of illustration.

Fig.~\ref{fig:Lzdotm2} shows the radial variation of torque in a pure 
exponential disc, in disc with stars and gas, and in our full galaxy model, i.e., 
stars+gas+dark matter halo due to a tightly wrapped $m=2$ perturbation. 
In all cases, the $m=2$ perturbations (e.g., spiral arms) transfers angular 
momentum outwards when they are trailing ($k > 0$), a result well-known 
for the last four decades \citep{LBK1972} and very recently ALMA reveals
the evidence of a trailing spiral carrying angular momentum outward in NGC 1566
 \citep{Combesetal2014}. 
It is interesting to note that in the long-wave limit, the net torque (gravity+advective), 
calculated in our full model of stars+gas+dark matter,
 beats the gravity torque alone by a factor of $\sim 3$ depending on the amplitude of 
the $m=2$ perturbation (a fact that had been suspected in the original paper by \cite{LBK1972}). 
For stars alone disc, the magnitude of the advective torque is roughly a factor of $2$ 
times more than the corresponding gravity torque. It is also worth to note that the
advective torque supports the gravity torque in the presence of an $m=2$ trailing
spiral.

\section{Slowly rotating lopsided pattern}
\label{sec:slow}

The pattern speed of a lopsided perturbation is an unknown quantity as
there is no observational measurement attempted so far unlike the galactic
bar \citep{Jog2011}. In its absence, we assume a slowly rotating lopsided
pattern imposed on the disc with $\Omega_{p} << \Omega$. An exponential 
stellar disc is shown to support lopsidedness as a discrete normal mode with such 
a small pattern speed and this has been further supported by N-body simulations
\citep[see][]{Sahaetal2007}. However, for $m=1$, one gets 
a peculiar situation, namely that if the pattern speed $\Omega_p > 0$, then there is no 
ILR (inner Lindblad resonance) as $\Omega -\kappa$ is negative, in general, for galaxies. 
On the other hand, an ILR can only exist for $\Omega_p < 0$ i.e., for a retrograde lopsided
perturbation. But then there can be no corotation \citep{Jog2011} which might not turn
out to be favourable for the lopsided perturbation. A galaxy can evolve without an ILR; 
in fact for certain non-axisymmetric perturbation (e.g., a bar or even lopsidedness) it 
can be a bonus. An absence of a strong ILR is desirable to complete the feedback loop 
required for the swing amplification \citep{Toomre1981} through which these perturbation 
might actually grow. 
So we consider a small prograde pattern speed for the lopsidedness so that 
the corotation is placed at very large radii in the disc.  

\begin{figure}
\rotatebox{270}{\includegraphics[height=8.5cm]{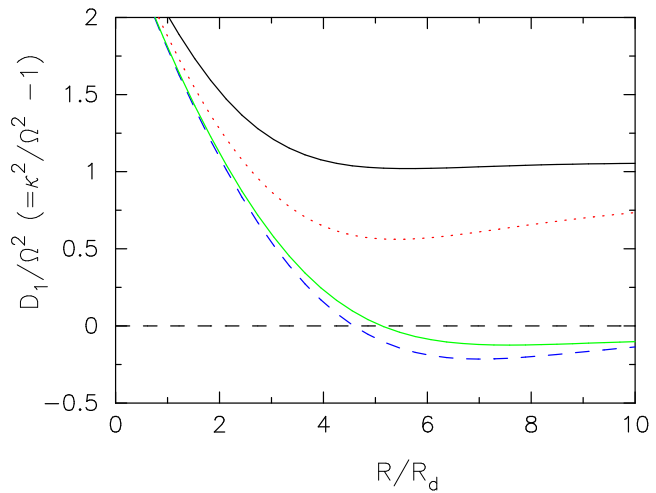}}
\caption{$D_1/\Omega^2$ parameter versus radius for an exponential disc with stars-alone, 
stars+gas and stars+gas+dark matter halo. The star-alone disc and the stars 
plus gas disc show an "unphysical" situation, namely a change of sign in 
D$_1/\Omega^2$ and hence the advective torque $C_{advec}$ due to $m=1$ perturbation 
(see subsequent figures). This problem is solved on addition of the dark matter halo 
where the mass builds up more gradually at large radii. The black solid and blue dashed
lines represent stars+gas+dark matter($\rho=0.035$) and stars+gas+dark matter ($\rho=0.01$)
respectively. }
\label{fig:D1}
\end{figure}

In this limit, we have $D_{1}\simeq \kappa^2 - \Omega^2$. In Fig.~\ref{fig:D1}, 
we show that the radial 
variation of this parameter $D_1$ which appears to play a pivotal role in deciding the 
direction of angular momentum transfer due to advective mechanism in galaxies. It is 
clear from the figure that for a pure exponential disc, the parameter
$D_1$ flips its sign from positive to negative at a radius of $4.6 R_d$, which 
we call as the {\it transition radius}, denoted as $R_{tr}$. This transition corresponds 
to when $\kappa$ the epicyclic frequency, falls below $\Omega$ the angular speed.
For a self-gravitating disc, this may seem counter-intuitive since for most realistic 
rotation curves $\kappa$ is $\geq \Omega$. However, such a situation can occur for 
example, for highly centrally concentrated (or super-Keplerian or steeper than Keplerian) 
mass distributions, including an exponential disc as shown here. Interestingly, a number 
of mass distributions such as the Toomre's model~n, or 
a Gaussian density distribution, which are widely used in modelling astrophysical discs 
also possess a transition radius ($R_{tr}$). These cases are given in the Appendix, 
and their implications are discussed in section~\ref{sec:discuss}.

\begin{figure}
\rotatebox{270}{\includegraphics[height=8.0 cm]{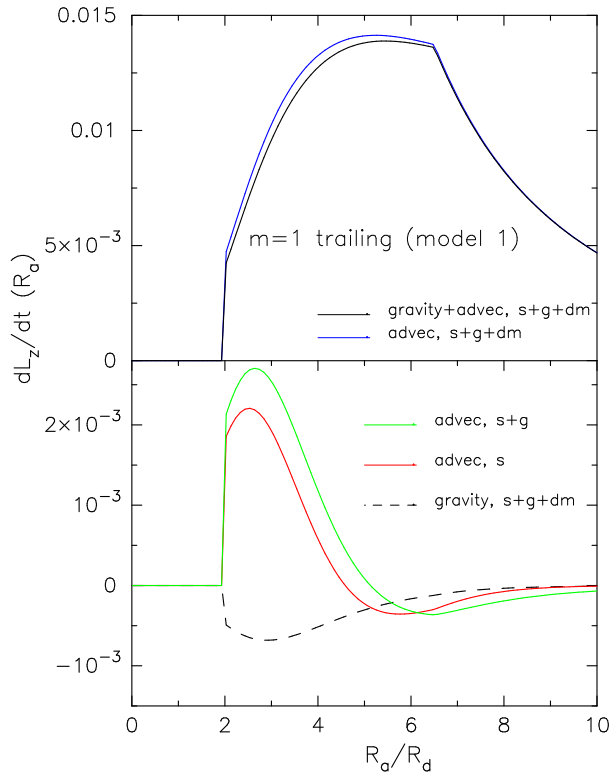}}
\caption{Radial variation of the torque i.e., the rate of change of angular momentum 
in the disc for an $m=1$ trailing lopsidedness (model~1, corresponding to left panel 
of Fig.~\ref{fig:surfA1A2} ). Note that in the presence of a dark halo, the disc 
gains angular momentum (by advective transport) for a trailing $m=1$ at all radii. 
The unit of $dL_z/dt$ is given by $G M_{d}^2/R_d$. }
\label{fig:Lzdotm1j}
\end{figure}

In Fig.~\ref{fig:Lzdotm1j} and Fig.~\ref{fig:Lzdotm1}, we show the angular momentum transport
by the $m=1$ lopsided perturbations as shown in Fig.~\ref{fig:surfA1A2}, first, for an 
exponential disc distribution. As can be seen from these figures, the gravity torque alone 
is negative at all radii when the $m=1$ lopsidedness is trailing. It is interesting to note 
that the advective torque is dominating the gravity torque roughly by a factor of $2$ and 
oppose the gravity torque (unlike the $m=2$ trailing case). However, sign of the 
net torque (advective+gravity) flips exactly at the transition radius of the exponential 
disc. The presence of a transition point hampers the smooth angular momentum flow in 
the disc -- irrespective of whether it is trailing of leading. This, in turn, might lead 
to an unfavourable situation for the evolution of galaxies. However, such a situation 
can only be completely eliminated if we add sufficient amount of gas in it and/or embedded 
the disc in a dark matter halo as discussed below.      

\subsection {Inclusion of gas and dark matter halo}
\label{sec:gas}

If we include a reasonable amount of cold gas in the disc (say 10-15\% )- as seen 
in late-type galaxies \citep[e.g.,][]{BT1987}, then the 
transition radius, $R_{tr}$ changes from 4.6$R_d$ to about 5.1 $R_d$ (see Fig.~\ref{fig:D1}). 
This is not because of any physical property such as dissipation related to the gas 
component per se but because of the gas distribution that is taken to fall slower 
than that of stellar distribution here as observed (see Eq.~2, Section~\ref{sec:model}).
If we were to add more gas, say more than 20\% by disc mass, then we do not have 
the transition radius any more in the disc. But gas fraction higher than 20\% 
is not likely since that would make a disc highly unstable to the growth of 
gravitational instabilities \citep{Jog1996}.  

The transition point is also avoided when the disc is embedded in a dark matter halo. 
In either case, the net $\kappa$ exceeds $\Omega$ and hence the transition point is 
avoided (see Fig.~\ref{fig:D1}). In the presence of a dark matter halo that gives 
rise to a flat rotation curve, we get $D_1 > 0$ at all radii (where $V_c$ is constant 
correspnding to $\rho_{0}=0.035 M_{\odot}pc^{-3}$);
the same holds true even when the rotation curve is moderately falling (see 
Fig~\ref{fig:kinematiclop} and Fig.~\ref{fig:D1}) which arises when use the other dark halo model with central density 
$\rho_{0} = 0.01 M_{\odot}pc^{-3}$. Adding some amount of gas makes $D_1/\Omega^2$ to be 
slightly higher than 1 (for the case with $V_c$ constant). These findings 
have significance for the slow evolution of galaxies as discussed next. 

\begin{figure}
\rotatebox{270}{\includegraphics[height=8.0 cm]{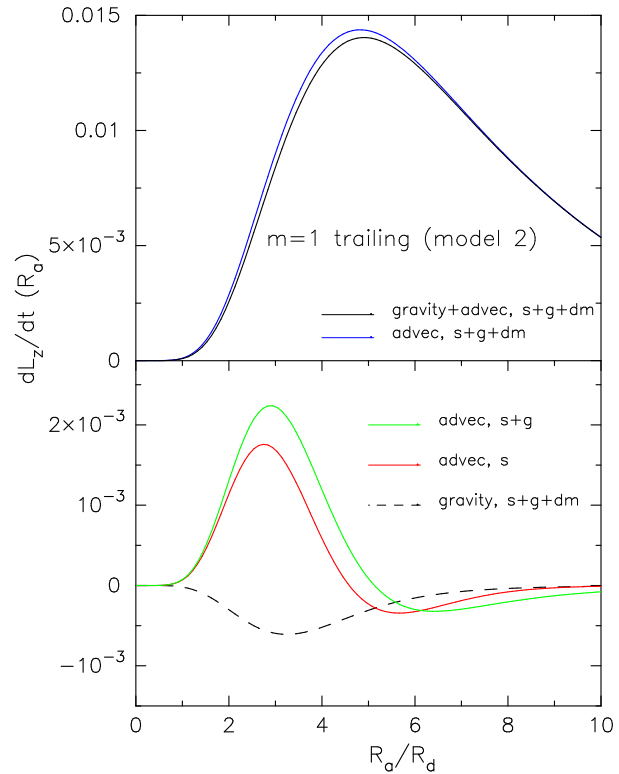}}
\caption{Same as in Fig.~\ref{fig:Lzdotm1j} but for model~2 corresponding to the
middle panel of Fig.~\ref{fig:surfA1A2}. Again due to advective transport, the disc 
gains angular momentum at all radii in the presence of a dark halo. The unit of $dL_z/dt$ is given by $G M_{d}^2/R_d$. }
\label{fig:Lzdotm1}
\end{figure}

The net transfer of angular momentum due to an $m=1$ trailing lopsided pattern
in an exponential stellar disc with gas and our full model with the dark matter halo 
(with $\rho_{0} = 0.035 M_{\odot}pc^{-3}$) having a flat rotation curve are shown 
in Fig.~\ref{fig:Lzdotm1j} and Fig.~\ref{fig:Lzdotm1}. The lopsided perturbation 
used in deriving these figures are the same as in the first two panels of 
Fig.~\ref{fig:surfA1A2}. As discussed below, the basic facts of angular momentum 
transfer do not change appreciably whether we use either models of $A_{1}(R)$.
Both these figures indicate that
the net transfer of angular momentum in the presence of a dark matter halo giving rise to 
a flat rotation curve is enhanced roughly by a factor of $7$ over that due to stars+gas alone. 
Note that a dark matter halo explicitly affects the advective torque by changing the 
free precession frequency ($\Omega_{pf}$) of the slow lopsided perturbation. 
Decreasing the amount of 
dark matter in the galaxy reduces the value of $D_1$ (see Fig.~\ref{fig:D1}) and hence 
the net torque also reduces. In other words, in a disc galaxy with a falling rotation 
curve (in our case, one with $\rho_{0} = 0.01 M_{\odot}pc^{-3}$), the net outward flow 
of angular momentum will be slower (approximately by a factor of 3 compared 
to the other dark halo producing the flat rotation curve) and the process will 
be slowest or inefficient in the limit where dark matter halo is totally absent. 
In such an extreme case, the disc alone is barely stable against local perturbations 
and a non-responsive dark matter halo is shown to be crucial to ensure local disc 
stability \citep{Jog2014} as well as to prevent global bar-like 
instability \citep{OstrikerPeebles1973}.    
The inclusion of a rigid dark matter halo in our picture has the effect that it removes the
transition point in the galaxy and the total torque is positive at all radii for the
trailing lopsided perturbations we consider here i.e., the disc within $R=R_a$ 
gains angular momentum. In other words, the inclusion of
a reasonable dark matter halo reinforces the angular momentum to flow inwards if the lopsidedness
is trailing. If this is allowed, such perturbation can not last long and can not drive 
the galaxy evolution.
On the other hand, if galaxies are to evolve by maximizing entropy, the angular momentum 
has to be transferred outward -- demanding that {\it lopsidedness has to be leading in nature.}

\begin{figure}
\rotatebox{270}{\includegraphics[height=8.5 cm]{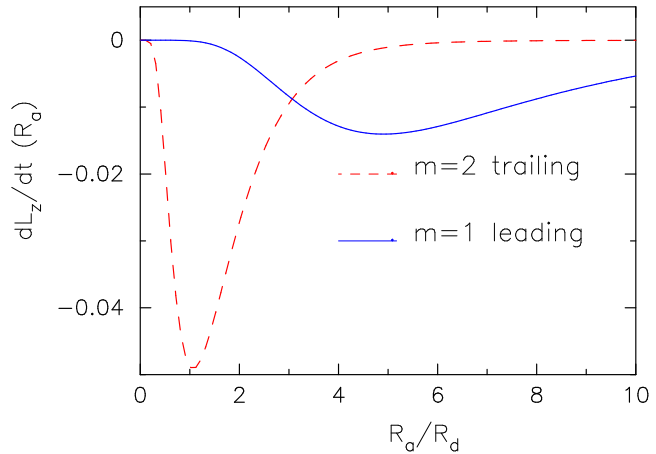}}
\caption{The plot showing the radial variation of the total torque (sum of gravity 
and advective) i.e., the net rate of change of angular momentum due to $m=1$ and $m=2$ 
perturbations for stars+gas+dark matter halo ($\rho_0 =0.035 M_{\odot}pc^{-3}$). 
Note that only $m=1$ leading perturbation can transport angular momentum
outward (i.e., the disc with $R_{a}$ looses angular momentum) opposite to the $m=2$
perturbation which has to be trailing to do so. The unit of $dL_z/dt$ is given by $G M_{d}^2/R_d$.}
\label{fig:Lzdotm1m2}
\end{figure}

In Fig.~\ref{fig:Lzdotm1m2}, we show that indeed it is the leading lopsided perturbation
which can drive the angular momentum flow outwards as the trailing $m=2$ perturbation does, 
albeit less vigorously in the inner regions. Note that for $m=2$, the final peak of the 
net AMT occurs approximately at a region where $A_2 (R)$ has its maximum. This can be 
understood as follows: the second
term for $m=2$ in the square bracket in Eq.[27] is nearly flat in the inner few scale 
length followed by a gentle rise in the outer parts of the disc where $A_2$ falls sharply 
to zero -- when multilied the net peak is basically determined by the $A_2$. For $m=1$, 
the second term is a smoothly rising function right from the center and $A_1$ is also 
rising but reaching its peak after $6$~$R_d$  (see Fig.~\ref{fig:modelA1})-- when multiplied 
the net peak occurs 
before $6$~$R_d$. In essence, the peak of AMT is basically determined by the radial variation
of the perturbation and the second term in square bracket in eq.~(27) which largely determines
the advective transport. 
For the kind of perturbations 
(Fig.~\ref{fig:surfA1A2}) we have used, the net torque due to the $m=2$ trailing perturbation
is an order of magnitude higher than $m=1$ leading perturbation within $2$ disc scale length. 
On the other hand, in the region beyond $\sim 4 - 6$~scale length, the torque due to the $m=1$
leading perturbation dominates and about $15$ times higher than that due to $m=2$ trailing wave.
It is trivial to see that by increasing the peak amplitude of $A_1$ by a factor of 2, we 
would increase $d {L}_{z}/d t$ roughly by a factor of 4 (see Eq. [27]). While the peak 
strength of the torque due to $m=1$ (seen in the range $4 - 6 ~R_d$) is a few times smaller
than due to $m=2$ (seen in $< 2 ~ R_d$) (see Fig.~\ref{fig:Lzdotm1m2}), the crucial point
is that in the outer parts, it is the $m=1$ perturbation that provides the channel that leads
to the ourward AMT.

\vspace{-0.5cm}
\section{Discussion}
\label{sec:discuss}

\noindent {\bf 1. Relative roles of $m=1$ and $m=2$ in AMT}

The structure of a spiral galaxy is such that the inner region is dominated 
either by a bar and/or spiral arms and generally the outer part is dominated 
by lopsided asymmetry. There might be a deeper reason for such a configuration
to exist in nature. Following \cite{LBK1972}, and a large amount of research by a number of 
authors  \citep{TremaineWeinberg1984, Weinberg1985, HernquistWeinberg1992, Zhang1999,
WeinbergKatz2002,Athanassoula2002,Athanassoula2003,SellwoodDebattista2006,Dubinskietal2009,
Sahaetal2010,Sahaetal2012}, the important role played by a bar and spiral arms in 
galaxies has now been understood. Both a bar and spiral arms help in evolving an 
initially axisymmetric disc galaxy by redistributing energy and angular momentum so 
that the disc reaches a state of minimum energy configuration. The end product of 
this process is a growing central concentration or a pseudo-bulge formation 
\citep{CombesSanders1981,PfennigerNorman1990, Rahaetal1991, KormendyKennicut2004,SahaNaab2013}.

However, often this process requires gas to flow inwards. Plenty of 
observations suggest that galaxy accretes gas from outside
either as a smooth infall - as shown by extended gas filaments as in NGC 891 
\citep{Mapellietal2008}, or via satellites \citep[e.g.,][]{ZaritskyRix1997}, also see
the recent review on gas accretion by \cite{Combes2014}. 
Numerical simulations also indicate that galaxies accrete gas along cosmological 
filaments \citep{Keresetal2005, Dekeletal2009}. The importance of both $m=1$ and $m=2$
modes in driving gas inflow has been stressed previously by \cite{Combes2001}. But details
about the angular momentum transport by $m=1$ has not been worked out. As we understand now, 
unless there is a positive definite torque acting on the gas, it is hard for an inflow 
to occur in the first place or there has to be other mechanisms at work.
We show that lopsidedness can act as a bridge between the stellar disc and the 
cosmic filaments which are a reservoir of cold gas. This galactic leading $m=1$ mode is 
a machinery behind bringing gas to the edge of the stellar disc from where the 
gas inflow is taken care of by the spiral arms and bars. This is in compliance with
the fact that the strength of lopsidedness decreases as one move inward towards 
the galactic center where generally $m=2$ dominates.

We show using arguments similar to Lindblad, that on kinematic grounds, the outer parts 
of an exponential disc are susceptible to an $m=1$ mode. 
However, this has a transition point which marks the change of sign of AMT in the disc.
This is removed if the disc is embedded in a dark matter halo so that the net 
$\kappa$ dominates over $\Omega$ at all radii and an outward flow of angular momentum 
occurs smoothly at all radii, if the lopsided mode is leading. This gives a deeper 
meaning to the role played by the ubiquitous $m=1$ modes in shaping or restructuring galaxies.

Thus, the $m=1$ mode bridges the gap between the gas accretion from filaments onto 
the outer galaxy to transport to the inner galaxy. Inside the optical disc, the spiral 
arms take over as the main driver of AMT. Thus $m=1$ in the outer parts and $m=2$ in 
the inner parts together allow the secular galaxy evolution to proceed in a meaningfull way. 
The $m=1$ leading and $m=2$ trailing modes can be thought of as collaborators in a 
relay process where the $m=1$ mode facilitates gas infall up to about half the optical 
radius, inside of which the $m=2$ takes over as the agent that allows gas infall. 
Thus together these 
modes cause an outward AMT and allow a smooth secular evolution of a typical galactic 
disc. This then is the deep physical reason for the existence of $m=1$ mode, analogous 
to what \cite{Lynden-Bell1979} had shown $m=2$ to be a mechanism that structures spirals 
within the inner optical disc.

\noindent {\bf{2. Discs in dark matter halo vs. Bare discs}}

Interestingly, the other typical disc mass distributions such Toomre's model~$n=2$ 
(as used e.g., by \cite{Bournaudetal2005} and others), and a Gaussian distribution 
as seen in the HI in some galaxies \citep{Angirasetal2006, vanEymerenetal2011a} also 
require a dark matter halo to have a smooth flow of angular momentum 
(see the Appendix for details). 
Surprisingly, the power-law mass distributions that are often seen as bare systems 
as in pre-stellar regions \citep{LizanoShu1989} do not exhibit such a transition region 
and hence do not have a discontinuity in the outward AMT, which may explain why such 
bare systems can exist.
Conversely, an exponential or a Gaussian radial mass distribution cannot exist as bare 
discs without the cushioning of a dark matter halo that removes the transition point.

\section {Conclusions}
\label{sec:conclusion}
In this paper, we have studied the angular momentum transport mediated by a slowly rotating
lopsided asymmetry due to the standard or gravity torques as well as the lorry or advective 
torques. In particular, we work out the lorry transport of angular momentum in considerable 
detail to understand the role played by the ubiquitous lopsided perturbation in driving 
evolution in lopsided disk galaxies. Our main conclusions are the following:

We show that in the long-wave limit, the magnitude of advective torque due to a lopsided 
perturbation dominates over the gravity torque and opposes the angular momentum flow
due to the gravity torque unlike the case of an $m=2$ perturbation.

We show that in an exponential stellar disc with or without cold gas, there is a 
transition point at which a kinematic lopsided perturbation changes from retrograde to
prograde. We show that there are other mass distributions e.g., Toomre's model~n,
Gaussian distribution, for which this is also true. 
It is shown that in the presence of a transition point, the net angular momentum flow
due to a lopsided perturbation is hindered in such a mass distribution.

We then explicitly show that in the presence of a dark matter halo such a transition point 
is uplifted, allowing a smooth flow of angular momentum in the disc. 
In a typical lopsided galaxy, where an exponential disc is embedded in a dark matter halo,
an outward smooth transport of angular momentum occurs only when the lopsidedness is leading
in nature. This can facilitate smooth gas infall in the galaxy.

\section*{Acknowledgement}
The authors thank the anonymous referee for very constructive and insightful comments on
the manuscript.

\vspace{-0.5cm}
\appendix

\section{Application to other density distribution}

\subsection{Toomre's model~n}

The surface density distribution for Toomre's model~n is given by \citep{Toomre1963}

\begin{equation}
\Sigma_n(R) = \Sigma_0 (1 + \frac{R^2}{2 n R_T^2})^{-(n+1/2)},
\end{equation}

\noindent where $R_T$ is the scale length and $n$ an integer. Note that in the 
limit $n \rightarrow \infty$, the above density distribution
becomes a Gaussian density distribution which we discuss separately in the 
next section. 
The potential for this disc model can be obtained using the Bessel function expansion
approach as

\begin{equation}
\Phi_n(R,z) = \int_0^{\infty}{S_n(k)J_0(kR)e^{-k|z|} dk},
\end{equation}

\noindent where the function $S_n(k)$ is given by

\begin{equation}
S_n(k) = -2 \pi G \int_0^{\infty}{J_0(kR)\Sigma_n(R) R dR}
\end{equation}

Solving for the above integral, we have the potential for the Toomre's model~n 

\begin{equation}
\Phi_n(R,z) = -\Lambda_n \int_0^{\infty}{k^{n-1/2} J_0(kR) K_{n-1/2}(\alpha_n k) e^{-k|z|} dk}\end{equation}

\begin{figure}
\rotatebox{270}{\includegraphics[height=7.0 cm]{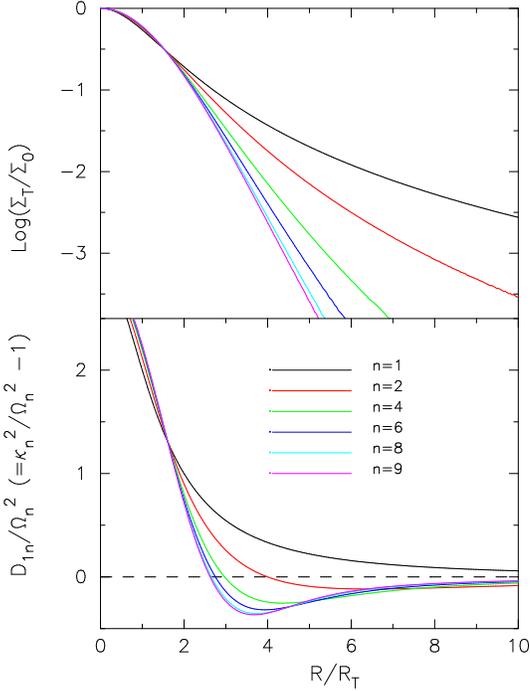}}
\caption{Surface density distribution and the corresponding $D_1$ parameter for the family
of Toomre's model~n. For n=1, it is actually the Kuzmin disc.}
\label{fig:toomren}
\end{figure}

In the above equation, 

\noindent $\Lambda_n = 4 \pi G \Sigma_0 R_T^2 \frac{n}{\Gamma(n+1/2)} (\alpha_n/2)^{n-1/2}$, $\alpha_n = 2 n R_T^2$ and $K_{n-1/2}$ denotes the modified Bessel function of 2nd order.

The azimuthal frequency for the disc can be derived using the following equation:

\begin{equation}
\Omega_n^2(R) = \Lambda_n \frac{2^n \Gamma(3/2) \Gamma(n+1)}{\sqrt{2}\alpha_n^{n+5/2}} F[n+1, \frac{3}{2};2; -\frac{R^2}{\alpha_n^2}]
\end{equation}

The ratio of radial epicyclic ($\kappa$) to azimuthal frequency can be obtained using
the following analytic expression:

\begin{equation}
\frac{\kappa_n^2}{\Omega_n^2} = 2[1 + \frac{F[n+1,\frac{3}{2},1;-\frac{r^2}{\alpha_n^2}]}{F[n+1,\frac{3}{2},2;-\frac{r^2}{\alpha_n^2}]}],
\end{equation}

\noindent where the functions $F$ refer to the hypergeometric function.
The $D_{1}$ parameter for the family of models are trivially given by

\begin{equation}
\frac{D_{1n}}{\Omega_{n}^2} = \frac{\kappa_n^2}{\Omega_n^2} -1 
\end{equation}

In Fig.~\ref{fig:toomren}, we show the surface density variation and the corresponding profiles
of $D_{1n}$. It is clear that all $n \ge 2$ Toomre's models~n have $\kappa/\Omega < 1.$ or 
$D_{1n} < 0.$ in some radial range. For higher $n$ models, the values of transition radii $R_{tr}$
decrease; although for really high $n > 8$,  $R_{tr}$ does not change appreciably. For $n=1$ 
Toomre model, there is no transition point so there is a smooth outward AMT. 

\subsection{Gaussian density distribution}

The surface density of a Gaussian disc is given by
\begin{equation}
\Sigma_{g}(R)= \Sigma_{g0} e^{-R^2/{2 R_g^2}}
\end{equation}

Again we determine the potential using the Bessel function approach outlined above:

\begin{equation}
\Psi_{g}(R,z=0) = -2 \pi G \int_{0}^{\infty}{dk J_{0}(k R) \int_{0}^{\infty}{dR^{\prime} R^{\prime} J_{0}(k R^{\prime}) \Sigma_{g}(R^{\prime})}},
\end{equation}

Solving this integral the potential for the disc distribution can be obtained as

\begin{equation}
\Psi_{g}(R,z=0) = -\sqrt{2 \pi} \pi G \Sigma_{g0} R_{g} e^{-y_{g}^2} I_{0}(y_{g}^2),
\end{equation}

\noindent where $y_g = R/{2 R_{g}}$.

It is then straightforward to derive the circular frequency in such a disc and is given by

\begin{equation}
\Omega_{g}^2(R) = \sqrt{\frac{\pi}{2}} \frac{\pi G \Sigma_{g0}}{R_{g}}e^{-y_{g}^2} \left[I_{0}(y_{g}^2) - I_{1}(y_{g}^2)\right]
\end{equation}

The ratio of $\kappa/\Omega$ for the Gaussian disc can be obtained from this formula:

\begin{equation}
\frac{\kappa_{g}^2}{\Omega_{g}^2} = 4 - \left[4 y_{g}^2 - \frac{2 I_{1}(y_{g}^2)}{I_{0}(y_{g}^2) - I_{1}(y_{g}^2)}\right]
\end{equation}

\begin{figure}
\rotatebox{270}{\includegraphics[height=7.0 cm]{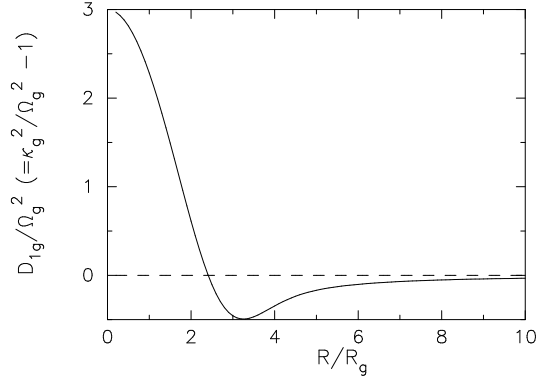}}
\caption{Radial variation of the $D_{1g}$ parameter for the Gaussian disc.}
\label{fig:kapaomggauss}
\end{figure}

In Fig.~\ref{fig:kapaomggauss}, we show the radial variation of $\frac{D_{1g}}{\Omega_{g}^2} = \frac{\kappa_{g}^2}{\Omega_{g}^2} -1$. The transition radius for this disc arises at $ R = 2.42 R_g$.

\subsection{Power law disc}

We consider the surface density distribution of a power law disc as

\begin{equation}
\Sigma_{p}(R) =  \Sigma_{\mathrm{p0}} (\frac{R}{R_{0}})^{-p}
\end{equation}

Using the Bessel function expansion approach as done in previous section, we have the
potential for a razor thin power law disc:

\begin{equation}
\Phi_p(R,z) = - 2 \pi G \Sigma_0 R_{0}^{p} 2^{-(p-1)} \frac{\Gamma(1-p/2)}{\Gamma(p/2)}\int_0^{\infty} {k^{p-2} J_0(k R) e^{-k{|z|}} dk}
\end{equation}

Which when computed at $z=0$ is given by

\begin{equation}
\Phi_p(R,0) = -\pi G \Sigma_0 R_0 K(p) (R/R_0)^{-(p-1)},
\end{equation}

\noindent where the function $K(p)$ is 

\begin{equation}
K(p) = \frac{\Gamma[1 - p/2] \Gamma[(p-1)/2]}{\Gamma[(3-p)/2] \Gamma[p/2]}  
\end{equation}

The allowed range for the power law exponent is $1 < p < 5/2$.

The circular frequency is given by

\begin{equation}
\Omega_{p}^{2}(R) = \pi G \Sigma_0 R_{0}^{p} (p-1) K(p) R^{-(p+1)} 
\end{equation}

The ratio $\kappa/\Omega$ for the power law discs turns out to be independent of 
radius r  and simply depends on the power law index:

\begin{equation}
\frac{\kappa^2}{\Omega^2} = 3 -p;  \: \: \: \: \: \: \: \: \: \:  1 < p< 5/2
\end{equation}

Thus, $\kappa/\Omega > 1$ for $p \leq 2$, hence there is no transition point, thus there is a smooth flow of angular momentum at all radii.

\end{document}